\documentclass[10pt]{article}

\usepackage{natbib}  
\usepackage[top=1in,bottom=1in,left=1in,right=1in]{geometry}  
\usepackage{bm}      
\usepackage{amssymb} 
\usepackage{amsmath} 
\usepackage{graphicx}  

\usepackage{amsthm}    

\theoremstyle{plain}

\newtheorem{corollary}{Corollary}

\newcommand{\pkg}[1]{{\fontseries{b}\selectfont #1}}   

\title{Pharmacokinetic Measurements in Dose Finding Model Guided by Escalation with Overdose Control}

\author{Arnab Kumar Maity$^{*}$, Satrajit Roy Chowdhury, Ray Li, Lada Markovtsova, Roberto Bugarini \\
$^{*}$arnab.maity@boehringer-ingelheim.com}

\date{22 May 2022}

\begin{document}
	
\maketitle

\abstract{Oncology drug development starts with a dose escalation phase to find the maximal tolerable dose (MTD). Dose limiting toxicity (DLT) is the primary endpoint for dose escalation phase. Traditionally, model-based dose escalation trial designs recommend a dose for escalation based on an assumed dose-DLT relationship. Pharmacokinetic (PK) data are often available but are currently only used by clinical teams in a subjective manner to aid decision making. Formal incorporation of PK data in dose-escalation models can make the decision process more efficient and lead to an increase in precision. In this talk we present a Bayesian joint modeling framework for incorporating PK data in Oncology dose escalation trials. This framework explores the dose-PK and PK-DLT relationships jointly for better model informed dose escalation decisions. Utility of the proposed model is demonstrated through a real-life case study along with simulation.}

\textbf{Keywords:} BLRM, DLT, MTD, Phase I, PK, Oncology.

\section{Introduction}  \label{section_introduction}

The early phase of a cancer clinical trial or an oncology clinical trial is unique by the virtue of phenomenon that the participants who are enrolled are not the healthy volunteers. The primary reason pertaining with this is severe toxicities generated as side effects of the investigating molecules which includes myelosuppression, immune suppression, nausea and vomiting, anorexia, alopecia and mucositis \citep{crowley2012handbook}. While a certain label of toxicity is acceptable and can not be avoided in order for the drug to show anti-tumor activities, severe toxicity is not desirable and this can be controlled by determining the appropriate dose of the investigating drug. Consequently, the primary objective of an early oncology trial is to find the highest dose which is well tolerated by the participants. This is often referred as the maximum tolerated dose (MTD) \citep{ji2010modified}. 

Several designs now exist to guide the dose escalation trial and each design has large literature associated with it. We refer the readers to \cite{geller1984design, rosenberger2002competing, hummel2009exploratory, zhou2018comparative} for a review of different methods in the literature. For a recent and comprehensive review see \cite{daimon2019dose}.  These designs can generally be classified in two categories -- conventional methods or rule based designs and model based designs \cite{rosenberger2002competing}. The conventional methods estimate MTD from the observed data while the latter approach sets the MTD as a parameter of the model and thereby tries to estimate from the data. 

The widely used rule based design is 3 + 3 design and there exist other variants of it. These are also referred as the trial and error method or up and down design due the nature of these designs. On the other hand, continuous reassessment method (CRM) \citep{o1990continual}, escalation with overdose control (EWOC) \citep{babb1998cancer}, and modified toxicity probability intervals (mTPI) \citep{ji2010modified} are prominent examples of the model based designs. It can be shown that CRM and EWOC are consistent in estimating the MTD \citep{o1996continual, tighiouart2010dose}. Our concentration in this paper centers around the model based design approach. In particular, in this article we develop our method to follow the EWOC principle. 

When applying EWOC criteria, it can be assumed that the dose toxicity relationship is logistic \citep{zacks1998optimal}. In other words, the probability that an individual will observe severe toxicity which would restrict the subject to receive the drug further is a logistic function of the dose. A dose finding model has generally been agreed to be adaptive such that it can accommodate all data until the current point of time and can drive the decision in an informative manner and Bayesian methodologies are particularly suitable for it. In addition, the non Bayesian parameter estimates often degenerate toward the boundary regions if there is no heterogeneity observed in the data which is often the case in the beginning of the trial \citep{o1996continual}. Toward this end, \cite{neuenschwander2008critical} proposed the Bayesian logistic regression model (BLRM) and provided some guidance on the prior elicitation. \cite{neuenschwander2015bayesian} further advocated and extended the mechanism for combination dose finding using the single agent studies as the prior data. 

The logistic regression model combined with EWOC criteria can be shown as optimal Bayesian feasible design a concept which was introduced by \cite{babb1998cancer} and will be illustrated further in Section \ref{section_method}. Furthermore, a desirable property of a good design is that the design would be coherent. \cite{tighiouart2010dose} proved that EWOC scheme satisfies the conditions of a coherent design. In fact, any relationship between dose and toxicity which is monotone can be shown to follow the coherent design properties which will be further illustrated in Section \ref{section_method}.

Historically it has been argued that any dose finding study has potential to become more effective in determining the MTD when the available pharmacokinetic (PK) data is considered in dose decision strategy \cite{collins1986potential, collins1990pharmacologically}. Since, the information of PK activity in vivo is an useful indicator in quantifying the drug amount applied, currently the Phase I Oncology studies use the PK data in a qualitative manner to aid the dose finding decisions, for example, see \cite{shaw2014ceritinib}. However, incorporating PK data in the dose response model has a potential to better estimate the MTD \citep{takeda2018bayesian}.

Nevertheless, attempts had been made to utilize the PK summary in developing the model based dose escalation approach. History dates back to \cite{collins1986potential} who suggested that a the MTD finding could be reached by considering smaller dose levels with a preclinically obtained pharmacokinatic guided design; since then different proposals of escalation designs involving PK data have been made in the literature; see \cite{collins1990pharmacologically, newell1994pharmacologically, holford1995target, piantadosi1996improved, whitehead2001easy, whitehead2007bayesian, cotterill2015practical, ursino2017dose, takeda2018bayesian} and references therein. An \textsf{R} \citep{R} package \pkg{dfpk} \citep{toumazi2018dfpk} have been developed to fit a PK guided Bayesian dose finding model. Furthermore, \cite{takeda2018bayesian} argued measuring the exposure data in terms of pharmacokinetics and taking them into consideration in finding the MTD is critical because there exist different dose exposure relationship such as linear and nonlinear, in different dose levels. Moreover, higher area under concentration (AUC) have been observed in those participants in the trial who experience dose limiting toxicity -- a dose which restricts participants to receive any further dose, than those who do not experience this event. Consequently, the thirst of incorporating PK data directly in the dose finding model is becoming more essential. 

\begin{figure}[h]
	\centering
	\includegraphics[height = 4 in, width = 4 in]{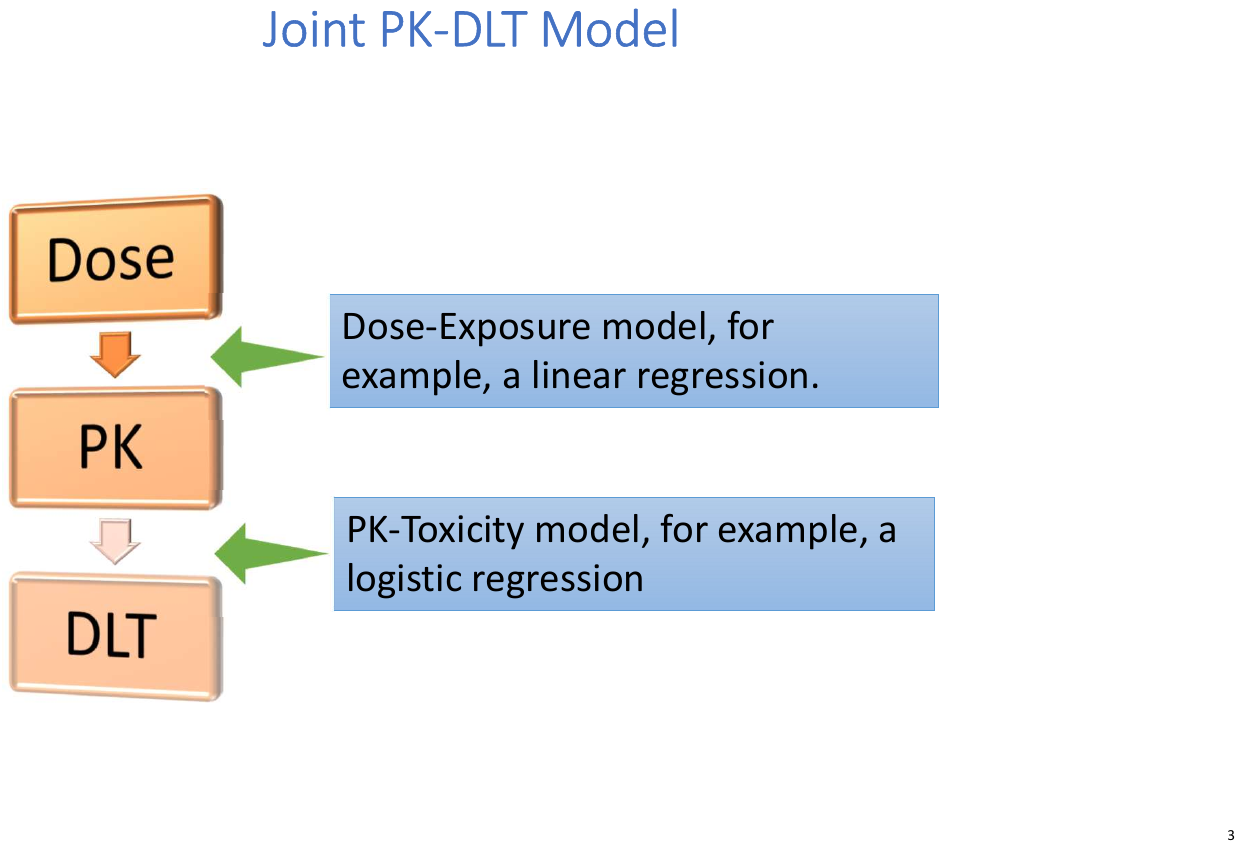}
	\label{figure_joint_model_schematic}
	\caption{A schematic representation of the proposed BLRM-PK joint model.}
\end{figure}

In this paper, we propose a shared parameter joint model approach which relies on EWOC principle and is presented in Figure \ref{figure_joint_model_schematic} schematically. The joint parameter consists of two components -- first model measures the dose exposure relationship and the second model quantifies the probability of experiencing a DLT as a function of PK exposure data. While we propose a Bayesian Log Normal regression to model the dose exposure relationship, the second model is fitted using a Bayesian logistic regression model and hence the inference is based on the usual EWOC criteria. The primary purpose of this approach is twofold -- first, the Bayesian logistic regression method has been shown to better estimate the entire dose toxicity curve \citep{cotterill2015practical} and one hopes to borrow the same feature here. Second, use of EWOC principle enables us to retain the current dose escalation practice which is straightforward and easy to implement using existing software routines such as East \citep{cytel1992east}. Moreover, it is shown that under mild conditions the proposed approach is coherent and consistent in MTD estimation. 

\cite{cotterill2015practical} proposed a similar two parts models in combining PK data in dual combination model, however, their model is not a joint model in a statistical sense and hence the consideration of PK data can only be utilized in setting the dose escalation rule and in setting the trial stopping rule. \cite{ursino2017dose} discussed various dose escalation models which incorporates PK data formally, however, their dose allocation rule is based on probabilistic computation. Conceptually, our proposed method is similar in spirit of the model formulation and the model fitting with that in \cite{takeda2018bayesian}. Even though they used the summary level PK data such as AUC which will be used in this article, they adjusted the departure of AUC values for a particular individual by considering a standardization over AUC measurements; in the contrary, we tackle this problem imposing a stochastic distribution assuming a Log Normal random variability. Furthermore, we retain the current practice of next safe dose declaration using EWOC principle. 

The rest of the article is organized as follows. Section \ref{section_method} formalizes the model and the method, discusses the parameter estimations, describes some theoretical properties of our proposed approach. In Section \ref{section_real_data} we apply our proposed methodology in two Phase I trial examples. In Section \ref{section_simulation} we provide some simulation studies and show that incorporating PK data is beneficial in terms of dose allocation. Essentially, the simulation studies show that more participants receive the doses which belong to the target toxicity interval than the case when PK measurements are not considered in the model. Finally in Section \ref{section_conclusion} we provide the concluding remarks and outline some future developments.

\section{Methodology}  \label{section_method}

\subsection{Joint Model}  \label{section_model}

Suppose $ n $ subjects are dosed $ d_i, i = 1, \ldots, n $ at different stages of a dose finding trial. While the response DLT $ y_i $ is binary taking values 1 or 0 for an individual is measured, the overall objective is to estimate the MTD. Suppose the pharmacokinetic exposure data is summarized by $ x_i $ for the $i$-th individual. The following models then can be used to describe the dose exposure relationship. 

\begin{equation}  \label{equation_dose_exposure_1}
\mu_i = \log(\gamma_0) + \gamma_1 \log(d_i/d^*), i = 1, \ldots, n,
\end{equation}
where
\begin{equation}  \label{equation_log_normal_model}
x_i \sim LN (\mu_i, \sigma^2), i = 1, \ldots, n.
\end{equation}
where, $ d^* $ is a reference dose which is used for standardization purpose. A guidance is given how to choose the value of $ d^* $ in Section \ref{section_prior}. A random variable $ X $ is said to follow a Log Normal distribution $ LN(\mu, \sigma^2) $ if and only if $ \log(X) $ follows the Normal distribution $ N(\mu, \sigma^2) $. We assume Log Normal distribution here because from a practical point of view any summary of PK measurements has support on the positive part of the real line. In addition, logarithmic transformations are considered here to bring stability in the computations. 

The second part of the joint model borrows the widely popular BLRM method \citep{neuenschwander2015bayesian} idea to predict the probability of experiencing a DLT in a given subject given PK originated from the given dose. The usual logistic regression model suffices to quantify this probability, 
\begin{equation}    \label{equation_exposure_DLT_1}
\pi_i = \frac{\exp\{\log \alpha + \beta \log(x_i)\}}{1 + \exp\{\log \alpha + \beta \log(x_i)\}}, i = 1, \ldots, n,
\end{equation}
with $ \pi_i $ being the probability of experiencing DLT, that is, $ y_i $ follows a Binary distribution with probability $ \pi_i $. In essence, together the models (\ref{equation_dose_exposure_1}), (\ref{equation_log_normal_model}), and (\ref{equation_exposure_DLT_1}) jointly can be fitted to the accumulated data to infer dose escalation decision. 

Note that, this joint BLRM-PK model will be referred as the BLRM-PK model hereafter. In addition, for the sake of prior specification convenience, we work with the logarithmic transformations of the coefficient parameters $ \gamma_1 $ and $ \beta $. Then models (\ref{equation_dose_exposure_1}) and (\ref{equation_exposure_DLT_1}) become,
\begin{equation}  \label{equation_dose_exposure_2}
\mu_i = \log(\gamma_0) + \log(\gamma_1) \log(d_i/d^*), i = 1, \ldots, n,
\end{equation}
and
\begin{equation}    \label{equation_exposure_DLT_2}
\pi_i = \frac{\exp\{\log \alpha + \log(\beta) \log(x_i)\}}{1 + \exp\{\log \alpha + \log(\beta) \log(x_i)\}}, i = 1, \ldots, n,
\end{equation}
respectively. Working with logarithm of the coefficients also have the interpretations such as dose level increases the exposure increases inside of the body and hence probability of observing a DLT increases. If the trial operators think that this assumption may not be tenable then it can be accommodated accordingly.

\subsection{Prior Elicitation}  \label{section_prior}

It has generally been agreed that BLRM-PK model is more useful after a trial accumulates at least some amount of data (personal communication with clinical pharmacology scientists). Toward this end, in a Bayesian setting after the trial starts to observe data then the data should dictate the dose recommendation by outweighing the prior. In this sense, a noninformative prior could serve the purpose, however, is not recommended primarily for two reasons. First, under a noninformative prior the DLT rates take a U-shape when no data is considered in the analysis, which is not desirable. Second, the Bayesian MCMC often observes divergence under a noninformative prior with a high uncertainty, that is, with low precision. To this direction, a weakly informative prior has been suggested in \cite{neuenschwander2015bayesian}. We follow similar approach here in eliciting an appropriate to carry out the Bayesian analysis. 

We set a Bivariate Normal prior for $ (\log \alpha, \log, \beta) $ with mean (logit(0.33), 0), standard deviation (2, 1) and correlation coefficient 0, where $ \text{logit}(x) = \log(\frac{x}{1 - x}) $. In addition, another Bivariate Normal prior is specified for $ \log \gamma_0, \log \gamma_1 $ with mean (0, 0), standard deviation (2, 1), and correlation coefficient 0. To complete the prior analysis a Log Normal prior with mean 0.25 and standard deviation 0.35 can be specified on $ \sigma^2 $. We select a low value for the mean for convergence purpose and the standard deviation 0.35 is obtained by $ \log(2)/1.96 $. As it will be shown in Section \ref{section_real_data}, this prior specification provides a wide range coverage of the DLT rates on the interval (0, 1). It can also be shown that this prior elicitation provides the anticipated DLT rate 0.33 at the dose $ d^* $.

\subsection{Inference and Decisions}  \label{section_model_inference}

For a dose escalation (or de-escalation) decision at any given point of time during a clinical we follow the process described by \cite{neuenschwander2015bayesian}. Hence the DLT rates at any dose can be categorized into three subgroups implying two extreme intervals are not intended as they either reflect an ineffective dose or they reflect a toxic dose. Formally, it is said that $ i $-th participant is under dosed if $ \pi_i < 0.16 $ and the participant is over dosed if $ \pi_i > 0.33 $. As a consequence one wants the more participants receive doses having DLT rate in between 0.16 and 0.33. In other words, it is a good practice of any Phase I trial that the dose finding method would maximize the probability $ \pi_i \in (0.16, 0.33) $. As a result, a good dose escalation design would be tailored by a good matrix of target toxicity (TT) probability which often ranges between 0.60 and 0.80 \citep{neuenschwander2015bayesian}. As a byproduct, we can also compute the probability of under dosing (UD) and probability of overdosing (OD) which are targeted to be minimized in a good design. 

The posterior probability of target toxicity can be computed by integrating the posterior DLT rate over the target toxicity region with respect to the posterior distributions, that is, 
\[
\Pr(\pi_d \in (0.16, 0.33)|\bm{x}, \bm{y}, \alpha, \beta, \gamma_0, \gamma_1) = \int_{\pi_d \in (0.16, 0.33)} \Pr(\bm{\pi}|\bm{x}, \bm{y}, \alpha, \beta, \gamma_0, \gamma_1) d \bm{\pi} d{\alpha} d{\beta} d{\gamma_0} d{\gamma_1},
\]
where $ \Pr(\pi_d|.) $ is the posterior DLT rate at dose $ d $. 
 
In a Bayesian design these probabilities can be obtained from the Markov Chain Monte Carlo (MCMC) as the exact analytical form of the posterior distribution of $ \pi_i$ is not available and we use JAGS \citep{plummer2003jags} and \pkg{rjags} \citep{rjags}. In particular, the following approximation can be used to estimate the above integral:
\[
\Pr(\pi_d \in (0.16, 0.33) |\bm{x}, \bm{y}, \alpha, \beta, \gamma_0, \gamma_1) = \frac{1}{M} \sum_{m = 1}^M \Pr(\bm{\pi}^{(m)} |\bm{x}, \bm{y}, \alpha^{(m)}, \beta^{(m)}, \gamma_0^{(m)}, \gamma_1^{(m)}) I(\pi_d^{(m)} \in (0.16, 0.33) ),
\]
where $ M $ is the total number of MCMC samples and $ \alpha^{(m)}, \beta^{(m)}, \gamma_0^{(m)}, \gamma_1^{(m)} $ are the posterior estimates of $ \alpha, \beta, \gamma_0, \gamma_1 $ at $ m $-th iteration respectively. In a very similar fashion, one can compute the probability of underdosing and probability overdosing. 

Recall that the primary objective of the trial is to estimate the MTD, denoted by $ \xi $ hereafter, and we define the proportion of subjects receiving MTD as, 
\[
\theta = \Pr(y_i = 1|d = \xi), i = 1, \ldots, n.
\]

In our formulation, the sequence of dose levels can be shown to be Bayesian-feasible at level $(1 - \alpha)$ (see Section \ref{section_theory}). Following \cite{babb1998cancer} we set $ \alpha $ at 0.25 meaning that at any stage of the dose finding trial a next dose will be statistically recommended if the overdose probability at the next dose is less than 0.25. This fully defines the Bayesian adaptive and sequential dose escalation (or de-escalation) process which will be followed in this article.

\subsection{Statistical Properties}  \label{section_theory}

\subsubsection{Consistency}

In what follows we use similar notations as in \cite{tighiouart2010dose}. Specifically, we assume that there exist $ d^1 $ and $ d^2 $, $ d^1 < d^2 $, such that
\[
\Pr(y_i = 1|x_i, d_i = d^1) = 0, i = 1, \ldots, n,
\]
\[
\Pr(y_i = 1|x_i, d_i = d^2) = 1 - \epsilon, i = 1, \ldots, n,
\]
where $ 0 < \epsilon < 1 $ and $ \theta < 1 - \epsilon $. 

Moreover, we consider the dose toxicity relationship of the form 
\begin{equation}  \label{equation_dose_toxicity_relation}
\Pr(y_i = 1|x_i, d_i) = F\biggl(F^{-1}(1 - \epsilon) + \psi \bigl(\frac{d_i - d^1}{d^2 - d^1} \bigr) \biggr),
\end{equation}
where $ 0 < \psi^1 \le \psi \le \psi^2 $ for some $ \psi^1 $ and $ \psi^2 $. This model assumes that the quantiles of $ F $ are linear in the log-standarized dose $ z = \log[(d - d^1)/(d^2 - d)] $. In our setup $ F = F_1 * F_2 $, where $ F_1(x) = e^x/(1 + e^x) $, $ F_2(.) $ is an identity map, and $ F $ is a convolution of $ F_1 $ and $ F_2 $.

Now letting $ \phi = F^{-1}(1 - \epsilon) - F^{-1}(\theta) $ the MTD can be shown to be equal to $ -\phi/\psi $ on the log standardized scale. Moreover, let $ G(x) = F(F^{-1}(\theta) + \phi + x), g(x) G'(x), $ and $ \log d_1 = -\phi/\psi $ be the log dose administered to the first participant. Then for subject $ i $, log dose $ \log d_i \in L^* = [-\frac{\phi}{\psi^1}, -\frac{\phi}{\psi^2}]. $ Also let $ h(\psi) $ be a prior density function for the parameter $ \psi $ on $ [\psi^1, \psi^2] $. Then the sequence of dose levels $ {d_n} $ such that $ \Pr(\log d_n \le - \phi/\psi|\bm{x}, \bm{y}) \ge 1 - \alpha $ for all $ n \ge 2 $ is called Bayesian feasible at level $ (1 - \alpha ) $. Now, using Theorem 2.1 in \cite{tighiouart2010dose} it can be shown that the sequence of dose levels considering our formulation will be Bayesian-feasible at level $ (1 - \alpha) $.

\subsubsection{Coherence}

A desirable property of good design, primarily because of ethical concern, is that to allow a dose escalation if there is no DLT observed in the current dose level and that to allow a dose de-escalation if there is a DLT observed in the current dose level. This property is known as coherent property and \cite{cheung2005coherence} argued how a practical adaptive model based approach such as CRM is designed to satisfy this property. 

Suppose after reparameterization in terms of the MTD $ \xi $, $ F(d_i, \xi) = \Pr(y_i = 1|x_i, d_i) $ is the model given in (\ref{equation_dose_toxicity_relation}). The BLRM-PK method will be said to be coherent in escalation if $ d_n \ge d_{n - 1} $ whenever $ y_n-1 = 0 $, and it will be coherent in de-escalation if $ d_n \le d_{n - 1} $ whenever $ y_{n - 1} = 1 $ for all $ n = 2, 3, \ldots $. In essence, the BLRM-PK design will be coherent if it is coherent in both dose escalation and dose de-escalation. The following corollary then establishes that the joint BLRM-PK model preserves this coherence principle. 

\begin{corollary}
	If $ F(d, \xi) $ is nonincreasing in $ \xi $ for fixed dose $ d $, then the BLRM-PK method described in Sections \ref{section_model} and \ref{section_theory} is coherent. 
\end{corollary}

By virtue of constructions in (\ref{equation_dose_exposure_2}) and (\ref{equation_exposure_DLT_2}), $ F(d, \xi) $ is in fact nonincreasing in $ \xi $ for fixed dose $ d $ and hence the proof follows by similar argument in \cite{tighiouart2010dose}.

\section{Real Data Example}  \label{section_real_data}

In this Section we apply our proposed methodology in two real world data originated from two clinical trials and carry out the data analysis respectively.  

\subsection{Application 1: Phase I Trial (Exposure is an Increasing Function in Dose)}

\begin{figure}[h]
	\centering
	\includegraphics[height = 4 in, width = 4 in]{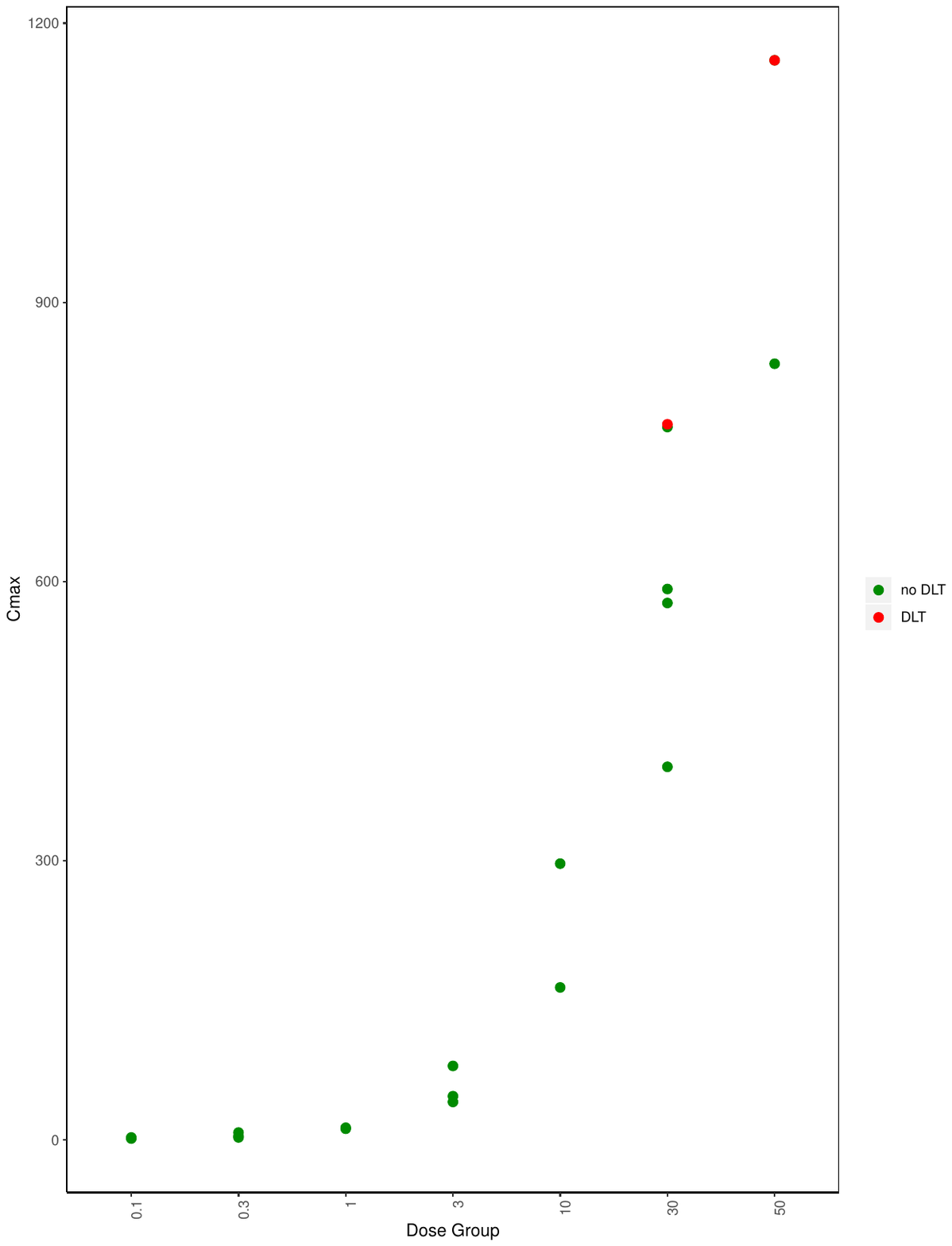}
	\label{figure_Cmax_App1}
	\caption{Cmax of the the participants in Example 1 have been plotted in the log scale. Participants who have experienced DLT (in red) have higher Cmax (exposure) values than those who have not experienced DLT (in green) for different administered doses.}
\end{figure}

This study enrolls 20 volunteers in the dose escalation part where the drug is given intravenously to the participants. The full data along with the DLT information and Cmax summaries are provided in Table \ref{table_dose_DLT_app1}. Cmax is computed by the clinical pharmacology scientists as a summary measure of the drug interactions with the body and is defined by the maximum serum concentration that a drug achieves in a specified compartment or test area of the body after the drug has been administered and before the administration of a second dose. Figure \ref{figure_Cmax_App1} provides the Camx measurements of the participants given the drug doses received by them. It can be noted that Cmax (exposure) is an approximately increasing function in the doses received by the subjects and the two DLTs (in red)  have occurred in the region that belongs to higher values of doses and higher values of Cmax.

\begin{table}[h]
	\centering
	\caption{Dose DLT Data of Application 1 -- The columns describe the dose given to the participants, DLT: 1 or 0 according to DLT is experienced or not, and Cmax summary.}
	\label{table_dose_DLT_app1}
	\begin{tabular}{crcrcrcr}
		\hline
		\hline
		Participant ID & Dose (unit) & DLT & Cmax & Participant ID & Dose (unit) & DLT & Cmax \\
		\hline
		\hline
		1  & 0.1  & 0 & 2.82  & 11 & 10.0 & 0 & 297.00 \\
		2  & 0.1  & 0 & 1.68  & 12 & 10.0 & 0 & 164.00 \\
		3  & 0.3  & 0 & 2.87  & 13 & 30.0 & 0 & 401.00 \\
		4  & 0.3  & 0 & 8.14  & 14 & 30.0 & 0 & 766.00 \\
		5  & 0.3  & 0 & 4.39  & 15 & 30.0 & 0 & 592.00 \\
		6  & 1.0  & 0 & 13.30 & 16 & 30.0 & 1 & 769.00 \\
		7  & 1.0  & 0 & 12.20 & 17 & 30.0 & 0 & 577.00 \\
		8  & 3.0  & 0 & 79.70 & 18 & 50.0 & 0 & 834.00 \\
		9  & 3.0  & 0 & 47.20 & 19 & 50.0 & 0 & 1160.00 \\
		10 & 3.0  & 0 & 41.00 & 20 & 50.0 & 1 & 1160.00 \\		
		\hline
		\hline
	\end{tabular}
\end{table}

\begin{figure}[h]
	\centering
	\includegraphics[height = 2 in, width = 2 in]{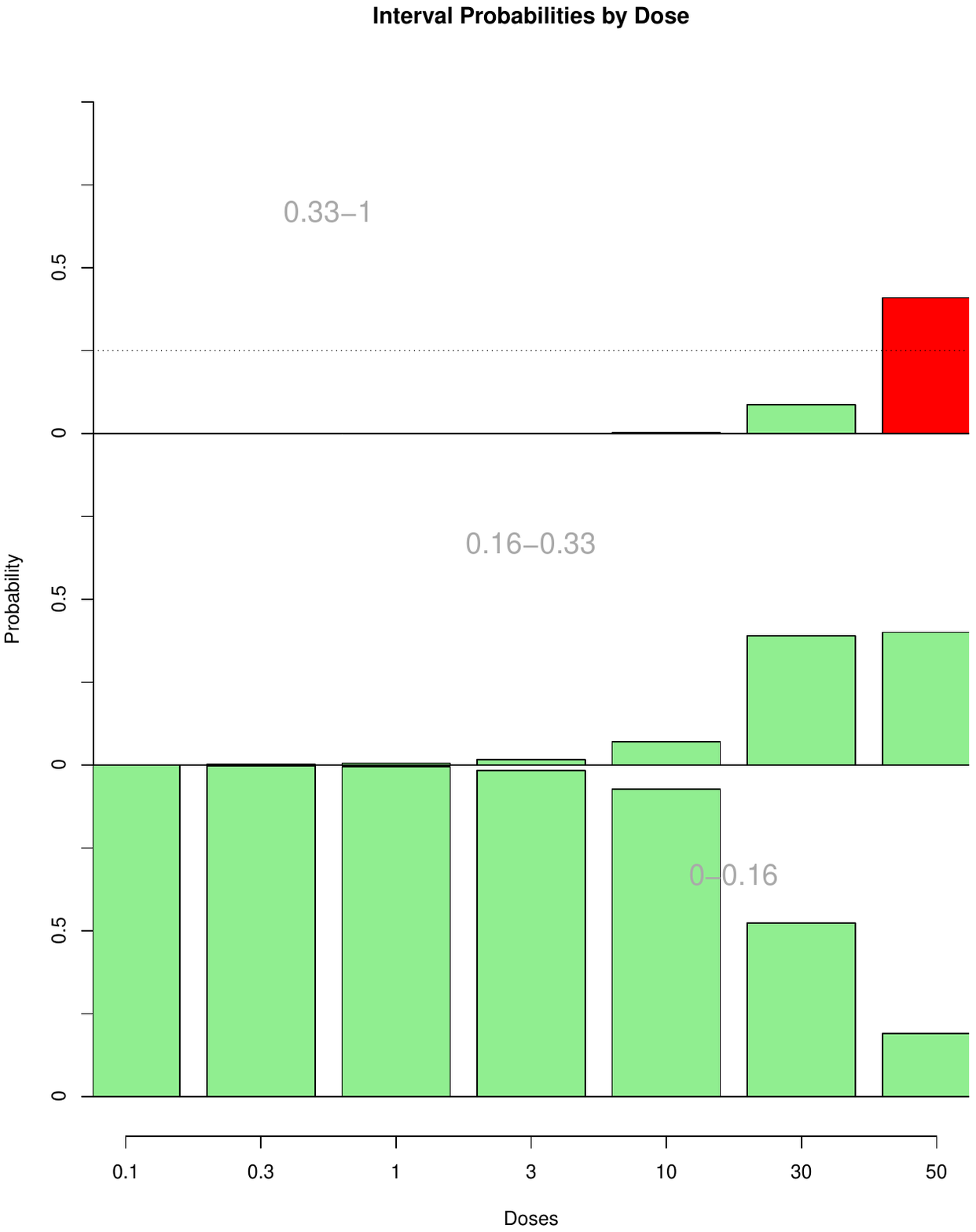}
	\includegraphics[height = 2 in, width = 2 in]{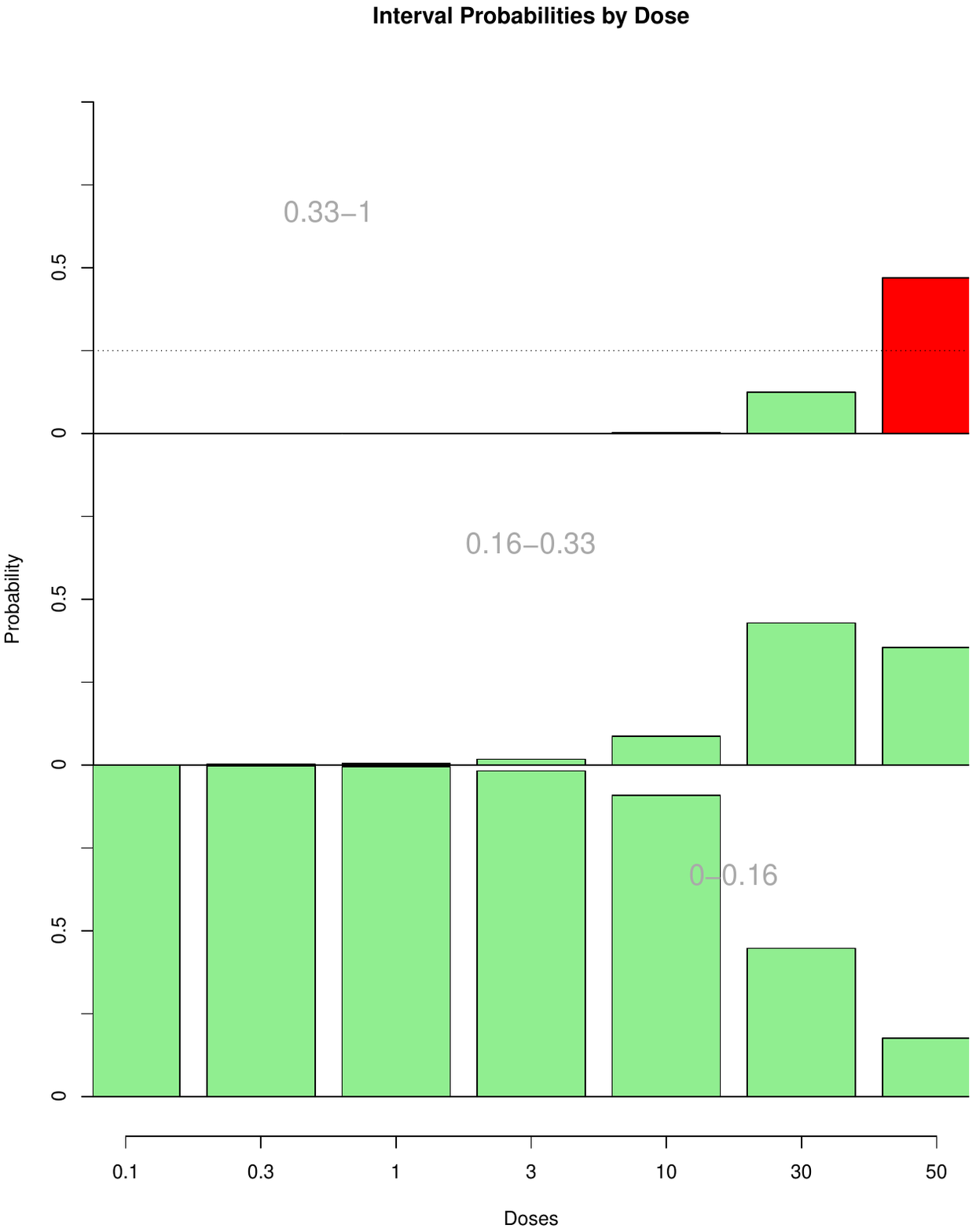}
	\label{figure_Posterior_DLT_Rate_App1}
	\caption{Posterior DLT rates after fitting BLRM (left hand side) and BLRM-PK (right hand side) in Example 1. The posterior probabilities are probability of underdosing (bottom panel), probability of target toxicity (TT), and probability of overdosing (OD). Dose that has more than 0.25 probability of overdosing is shown in red (EWOC criteria).}
\end{figure}

We fit both BLRM and BLRM-PK models to the data. To infer the safe dose recommendation we compute three probabilities -- probability of underdosing, probability of target toxicity, and probability of overdosing -- as discussed in Section \ref{section_model_inference}. To carry out the safe dose recommendation in optimize manner we follow the EWOC principle with $ \alpha = 0.25 $, that is, a dose will be said to be safe to receive if the overdose probability is less than 0.25. The resulting probabilities for each doses is plotted in Figure \ref{figure_Posterior_DLT_Rate_App1} after fitting BLRM (left hand side plot) and BLRM-PK model (right hand side plot). From the figures it is evident that dose 50 units can not be recommended as the next safe dose (in red) as the corresponding overdosing probability is beyond 0.25. This is confirmed by both the methods -- BLRM and the proposed BLRM-PK method. Intuitively, it means that higher the dose levels, the drug triggers higher Cmax in the body and thereby it increases the risk of toxicity observed by the volunteers.

\subsection{Application 2: Phase I Trial (Similar Exposure for Different Toxicity)}

The second application is a phase I clinical trial.
Even though the original study had different parts depending on different regimanes, we concentrate on the Phase I dose finding study which consists of the experiment where 39 study participants receives the investigational drug once in three weeks intravenously. The full dose information and dose limiting toxicity information is provided in Table \ref{table_dose_DLT_app2}. For the analysis we consider Cmax, a standard summary measurement in pharmaceutics, as in the previous example.

\begin{table}[h]
	\centering
	\caption{Dose DLT Data of Application 2 -- The columns describe the dose given to the participants, DLT: 1 or 0 according to DLT is experienced or not, and Cmax summary.}
	\label{table_dose_DLT_app2}
	\begin{tabular}{cccrcccr}
		\hline
		\hline
		Participant ID & Dose (unit) & DLT & Cmax & Participant ID & Dose (unit) & DLT & Cmax  \\
		\hline
		\hline
		1  & 0.13 & 0 & 5170   & 21 & 3.20 & 0 & 120000 \\
		2  & 0.13 & 0 & 4220   & 22 & 2.80 & 0 & 57800 \\
		3  & 0.33 & 0 & 8500   & 23 & 2.80 & 0 & 118000 \\
		4  & 0.33 & 0 & 14500  & 24 & 2.80 & 1 & 73900 \\
		5  & 0.83 & 0 & 29600  & 25 & 2.80 & 0 & 113000 \\
		6  & 1.40 & 0 & 71600  & 26 & 2.80 & 0 & 96000 \\
		7  & 1.40 & 0 & 60200  & 27 & 2.80 & 0 & 85200 \\
		8  & 1.87 & 0 & 78400  & 28 & 3.20 & 0 & 104000 \\
		9  & 2.47 & 0 & 102000 & 29 & 3.20 & 0 & 89600 \\
		10 & 2.47 & 0 & 115000 & 30 & 2.10 & 0 & 109000 \\
		11 & 0.83 & 1 & 20100  & 31 & 2.80 & 0 & 92100 \\
		12 & 1.87 & 0 & 50700  & 32 & 2.10 & 0 & 84100 \\
		13 & 2.47 & 0 & 110000 & 33 & 2.10 & 0 & 102000 \\
		14 & 1.40 & 1 & 49400  & 34 & 2.80 & 0 & 179000 \\
		15 & 1.87 & 0 & 31200  & 35 & 2.80 & 1 & 98200 \\
		16 & 2.47 & 0 & 40600  & 36 & 2.80 & 0 & 129000 \\
		17 & 1.40 & 0 & 87900  & 37 & 3.20 & 1 & 93300 \\
		18 & 1.87 & 0 & 81400  & 38 & 3.20 & 1 & 123000 \\
		19 & 2.47 & 0 & 101000 & 39 & 3.20 & 0 & 90300 \\
		20 & 2.47 & 0 & 149000 &    &      &   &       \\		
		\hline
		\hline
	\end{tabular}
\end{table}

\begin{figure}[h]
	\centering
	\includegraphics[height = 4 in, width = 4 in]{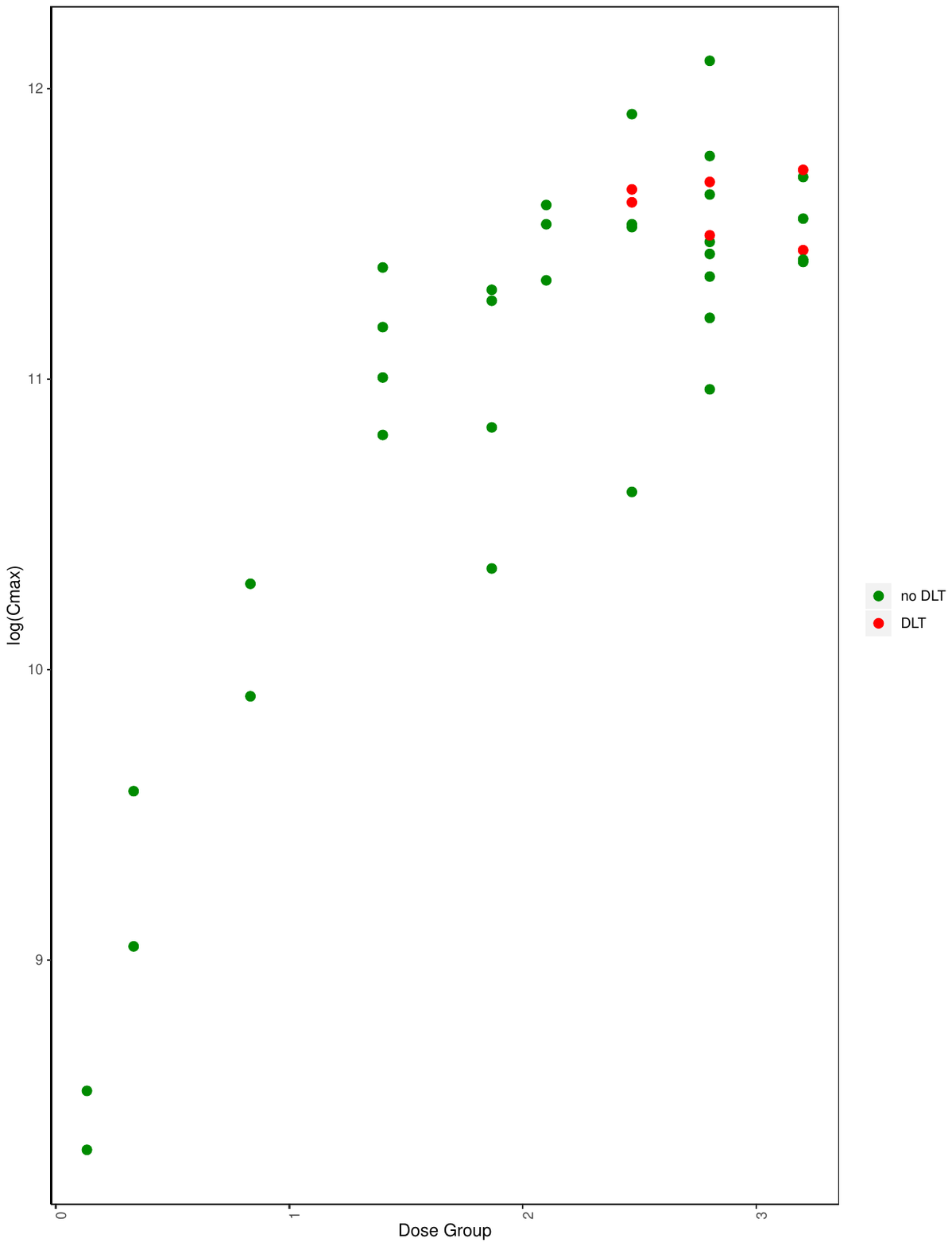}
	\label{figure_Cmax_App2}
	\caption{Cmax of the the participants in Example 2 have been plotted in the log scale. Participants who have experienced DLT (in red) and those who have not experienced DLT (in green) have similar Cmax (exposure) profile for different administered doses.}
\end{figure}

In Figure \ref{figure_Cmax_App2} we depict the observed Cmax of the participants across different doses administered on the subjects. It is interesting to note that the participants who have experienced DLT (in red) and those who have not experienced DLT (in green) have similar Cmax (exposure) profile for different administered doses. Hence, it would be worth to carry out the BLRM-PK analysis if the phramacokinatic data has any impact on the inference of the dose escalation decision when incorporating the pharmacokinetic data. 

\begin{figure}[h]
	\centering
	\includegraphics[height = 2 in, width = 2 in]{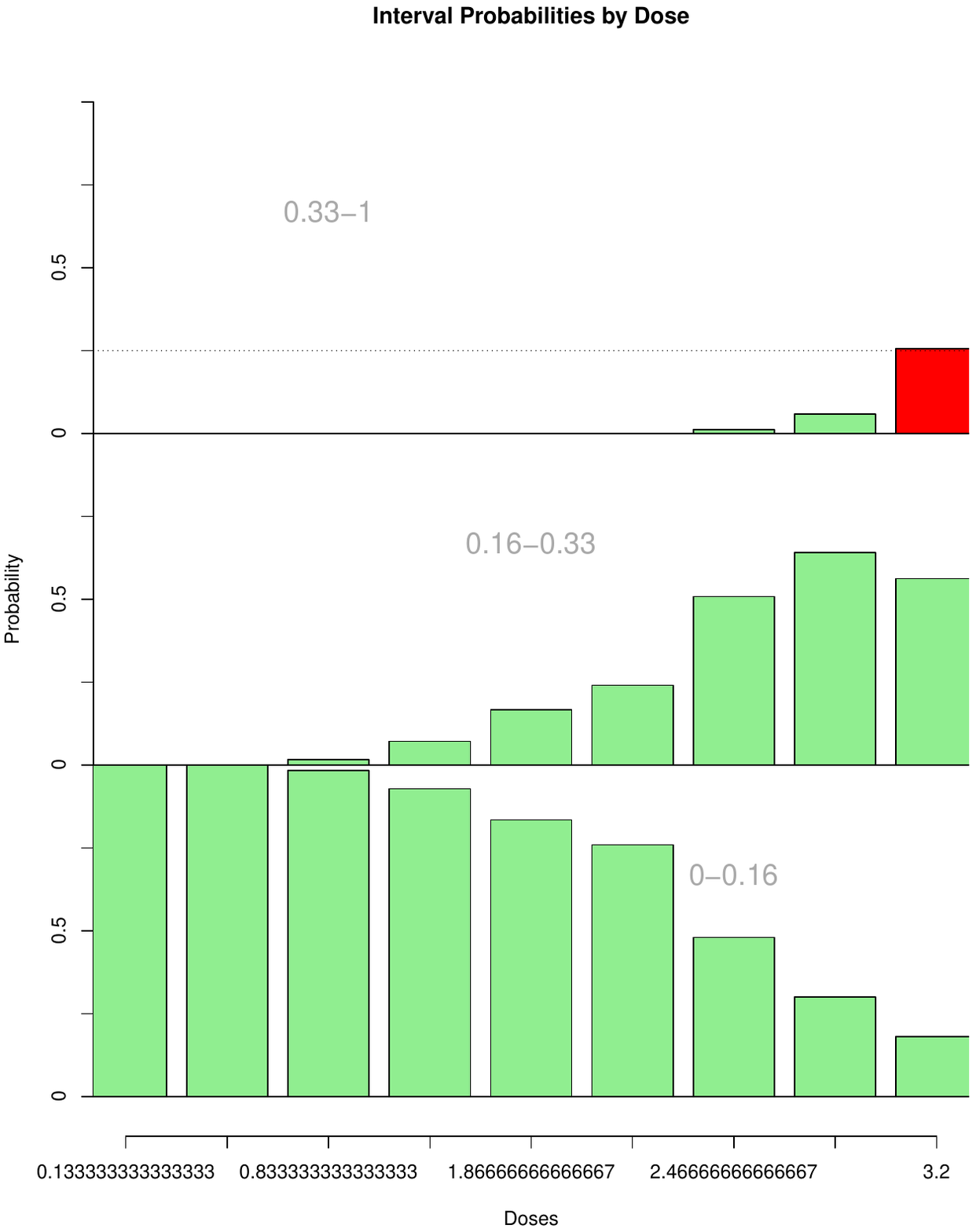}
	\includegraphics[height = 2 in, width = 2 in]{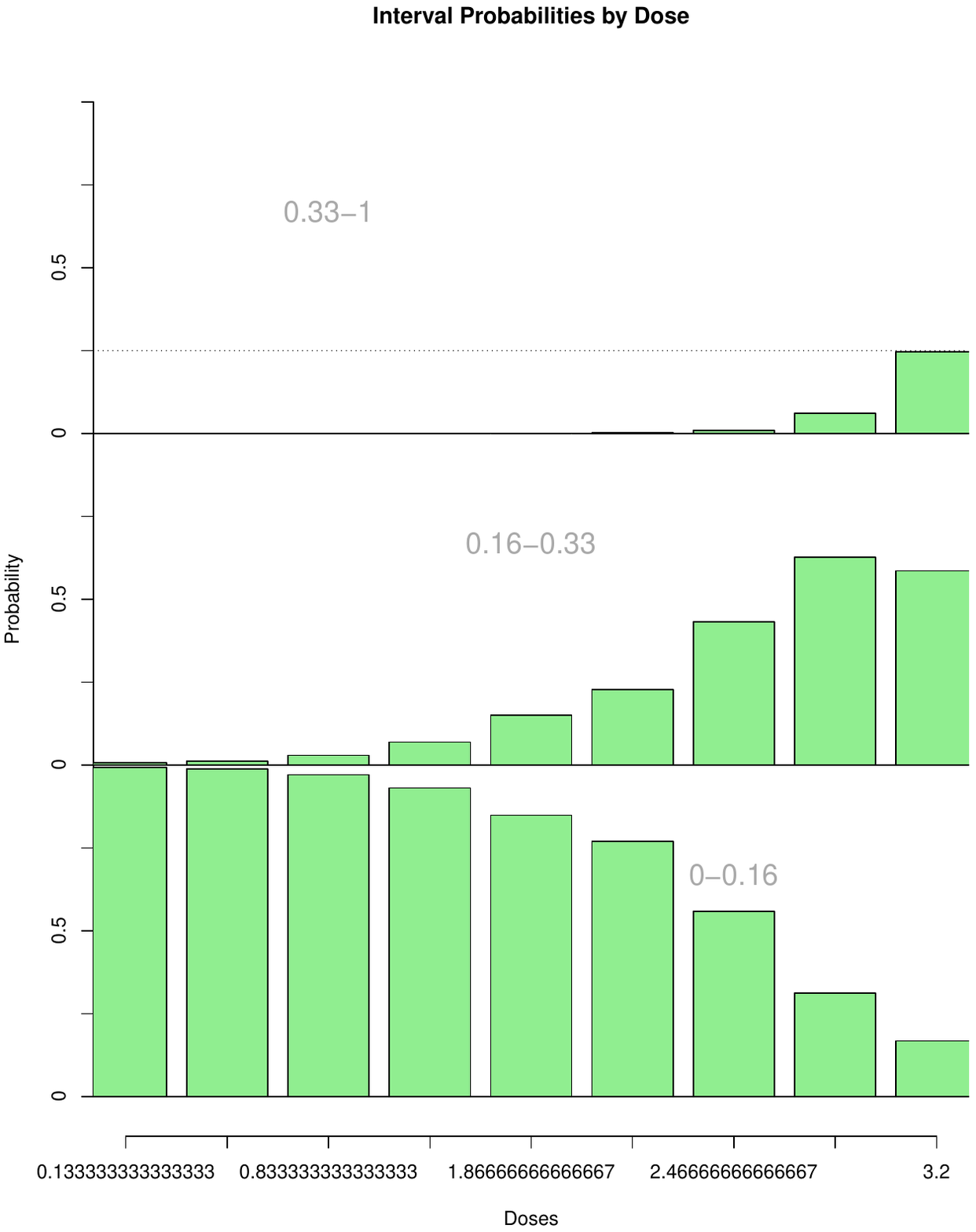}
	\label{figure_Posterior_DLT_Rate_App2}
	\caption{Posterior DLT rates after fitting BLRM (left hand side) and BLRM-PK (right hand side) in Example 2. The posterior probabilities are probability of underdosing (bottom panel), probability of target toxicity (TT), and probability of overdosing (OD). Dose that has more than 0.25 probability of overdosing is shown in red (EWOC criteria).}
\end{figure}

Toward this end, first the usual Bayesian logistic regression model (BLRM) is fitted to the data. The resulting posterior DLT rates can be summarized by left hand side plot in Figure \ref{figure_Posterior_DLT_Rate_App2} classified into probability of underdosing, probability of target toxicity, and probability of overdosing. According to the EWOC criteria described in Section \ref{section_method} a dose is said to be safe to administer at level $ \alpha $ if probability of overdosing is less than $ \alpha $. As in the previous example $ \alpha $ is set to 0.25. We note that the dose 3.2 units is not a safe dose according to this criteria. On the other hand, when the proposed BLRM-PK model is fitted in this data, the same EWOC criteria implies that dose 3.2 mg is a safe dose (Right panel of Figure \ref{figure_Posterior_DLT_Rate_App2}). Since the actual true scenario is not known in the real data, further study, for instance, a rigorous simulation study is needed to better understand the utility of the BLRM-PK method which is conducted in the next Section. 

\section{Simulation Study}  \label{section_simulation}

In this section we conduct a simulation study to illustrate the effectiveness of incorporating the pharmacokinetic information in the formal Statistical modeling via the proposed BLRM-PK model. We start with known exposure profile, see Table \ref{table_PK_toxicity} for each dose. In essence, at each dose we assume that it generates a given exposure according to Table \ref{table_PK_toxicity}. To be consistent with the model formulation in Section \ref{section_model} we assume the exposure data is in log scale in the sense that the actual exposure can be obtained by raising them in the power of exponential function. By setting $ (\log(\alpha), \log(\beta)) = (-2.5, 2) $ in (\ref{equation_dose_exposure_2}) the data generating toxicity profile is obtained as in Table \ref{table_PK_toxicity}.

\begin{table}[h]
	\centering
	\caption{Exposure and Toxicity Profile for Data Generation in Simulation.}
	\label{table_PK_toxicity}
	\begin{tabular}{rcc|rcc}
		\hline
		\hline
		Dose (unit) & Exposure & Toxicity & Dose (unit) & Exposure & Toxicity \\
		\hline
		\multicolumn{3}{c|}{First Scenario} & \multicolumn{3}{c}{Second Scenario}  \\ 
		\hline
		\hline
		0.1  & 0.40 & 0.15 & 0.13 & 0.05 & 0.125 \\
		0.3  & 0.47 & 0.17 & 0.33 & 0.13 & 0.133 \\
		1.0  & 0.53 & 0.19 & 0.83 & 0.28 & 0.151 \\
		3.0  & 0.60 & 0.21 & 1.40 & 0.67 & 0.210 \\
		10.0 & 0.67 & 0.24 & 1.87 & 0.75 & 0.223 \\
		30.0 & 0.73 & 0.26 & 2.10 & 1.10 & 0.289 \\
		50.0 & 0.80 & 0.29 & 2.47 & 1.15 & 0.299 \\
		     &      &      & 2.80 & 1.15 & 0.300 \\
		     &      &      & 3.2  & 1.16 & 0.302 \\
		\hline
		\hline
	\end{tabular}
\end{table}

The data generation and model fitting scheme proceeds in the following way. At dose level 0.1 we generate three DLT data points from a Binomial distribution with size 3 and probability 0.15. Moreover, we generate the pharmacokinetic summary from a log normal distribution with mean 0.40 and standard deviation 0.5. Then a BLRM is fitted using the dose DLT data and proposed BLRM-PK method is fitted using dose-DLT-pharmacokinatic data. Then the next safe dose is selected according to the EWOC criteria. The next dose recommendation is the maximum dose which satisfies the EWOC criteria but does not exceed maximum allowable increments. In this example the maximal allowable increment is set to 200\%, that is, the experiment can jump from the dose level 0.1 units to 0.3 units if EWOC criteria is satisfied. When the next dose is suggested by BLRM or BLRM-PK is selected, then again 3 data points are generated in a very similar manner as in first dose level. In this way the trial is carried out until the stopping criteria is met:
\begin{enumerate}
	\item No further escalation is possible.
	\item At least 6 patients have been treated at the recommended MTD $\overline{d}$.
	\item The dose $ \overline{d} $ satisfies one of the following conditions:
	\begin{itemize}
		\item The probability of target toxicity at dose $\overline{d}$ exceeds 0.5, i.e. $ \Pr(0.16 \le \pi_{\overline{d}} < 0.33) \ge 0.5 $;
		\item A minimum of 15 patients have been treated in the trial.
	\end{itemize}
\end{enumerate}

The maximum number of patients in dose escalation part of the trial is set to 50. We generate 1000 trials in this fashion and assess the performance of the proposed BLRM-PK method with that of BLRM. When assessing the performance of the methods, following matrices are populated and compared:
\begin{itemize}
	\item Percentage of participants receiving dose in the target toxicity interval
	\item Percentage of participants receiving an overdose
	\item Percentage of participants receiving an under dose
	\item Probability that recommended MTD at the end of the trial is in the target toxicity interval
	\item Probability that recommended MTD is an overdose
	\item Probability that recommended MTD is an under dose
	\item Percentage of trials with MTD below starting dose
	\item Average sample size.
\end{itemize}

\begin{table}
	\centering
	\caption{Simulation Result. TT: Target Toxicity, OD: Overdose, UD: Underdose, SD: Starting Dose.} 
	\label{table_simulation}
	\begin{tabular}{lrrrrrrrr}
		\hline 
		\hline
		Method & \multicolumn{3}{c}{Participant allocation} & $\Pr$(MTD & $\Pr$(MTD & $\Pr$(MTD & $\Pr$(MTD & Average   \\
		& in TT (\%) & in OD (\%) & in UD (\%) & in TT) & in OD) & in UD) &  is below & sample  \\
		& & & & & & & SD) & size \\
		\hline
		\hline
		\multicolumn{9}{c}{Simulation Example 1}  \\
		BRLM    & 51.1 & 0.0 & 48.9 & 0.69 & 0.0 & 0.02 & 0.29 & 10.0  \\
		BLRM-PK & 71.7 & 0.0 & 28.3 & 0.90 & 0.0 & 0.03 & 0.07 & 14.0  \\
		\multicolumn{9}{c}{Simulation Example 2}  \\
		BRLM    & 51.1 & 0.0 & 48.9 & 0.69 & 0.0 & 0.02 & 0.29 & 10.0  \\
		BLRM-PK & 69.2 & 0.0 & 30.8 & 0.86 & 0.0 & 0.05 & 0.09 & 13.8  \\
		\multicolumn{9}{c}{Simulation Example 3}  \\
		BRLM    & 25.3 & 0.0 & 74.7 & 0.41 & 0.0 & 0.28 & 0.31 & 15.9  \\
		BLRM-PK & 48.9 & 0.0 & 51.1 & 0.80 & 0.0 & 0.16 & 0.04 & 22.2  \\
		\multicolumn{9}{c}{Simulation Example 4}  \\
		BRLM    & 25.3 & 0.0 & 74.7 & 0.41 & 0.0 & 0.28 & 0.31 & 15.9  \\
		BLRM-PK & 45.3 & 0.0 & 54.6 & 0.74 & 0.0 & 0.21 & 0.05 & 21.5  \\
		\hline
		\hline
	\end{tabular}
\end{table}

We carry out four simulation examples and present the frequentist operating characteristics in Table \ref{table_simulation}. 
\begin{itemize}
	\item Simulation Example 1: PK measurements are generated from scenario 1 in Table \ref{table_PK_toxicity} with standard deviation 0.5.
	\item Simulation Example 2: PK measurements are generated from scenario 1 in Table \ref{table_PK_toxicity} with standard deviation 1.
	\item Simulation Example 3: PK measurements are generated from scenario 2 in Table \ref{table_PK_toxicity} with standard deviation 0.5.
	\item Simulation Example 4: PK measurements are generated from scenario 2 in Table \ref{table_PK_toxicity} with standard deviation 1.
\end{itemize}

From the numbers presented in Table \ref{table_simulation} it is evident that BLRM-PK method recovers the MTD which falls in the target toxicity region more than that of BLRM method. For instance, in case of Simulation Example 1, the estimated probability from the 1000 trials that the the dose identified by the MTD by BLRM-PK model is a dose from the target toxicity region is 0.90 while the same due to the BLRM method is 0.69. Furthermore, when BLRM-PK method is fitted then 71.1\% of the total participants receive a dose belonging to the target toxicity region while the BLRM method allocates 51.1\% of the total participants in the target toxicity region. Since it is the intend to maximize number of participants in the target toxicity region in an Oncology Phase I trial, obviuosly BLRM-PK method outperfoms the traditional BLRM method in this regard. We note that, the conclusion remains same across all the Simulation Examples.

\section{Conclusion}  \label{section_conclusion}

In this analysis we have shown that the joint modeling of dose-exposure and exposure-DLT approach has the potential to provide better understanding about the dose escalation curve. Furthermore it is more consistent design in estimating the correct MTD which is evident from the simulation. One assumption in the proposed model is that the dose exposure relationship is monotone which is a natural assumption according to the clinical pharmacology scientists (personal communication). Moreover, it allows to satisfy some sufficient conditions to consistently estimate the MTD. Additionally, by virtue of these assumption, it can be theoretically shown that the proposed design is coherent which is desirable property in dose escalation trials. In case, this assumption is tenable, then this can be relaxed and the model can be easily written to accommodate that. In our experience, we have seen that, in the long run, such as in simulation study this does not affect the conclusion about the superiority of the BLRM-PK approach. 

Our research is thus an important contribution to the overreaching goal of the ongoing efforts to model other important clinical measures and thereby informing the researchers quantitatively when a dose finding study is run. Nonetheless, it opens up the other avenues for other future research such as eliciting an appropriate prior for the dose exposure model parameters when and if any knowledge is available about the dose exposure relationship.

\bibliographystyle{apalike}
\bibliography{reference_blrmpk}
	
\end{document}